%% file: Paper-0019.tex
\documentclass[runningheads]{llncs}
\usepackage[T1]{fontenc}
\usepackage{graphicx}
\usepackage{booktabs}
\usepackage{hyperref}
\usepackage{tabularx,booktabs}
\usepackage{xcolor}
\usepackage{orcidlink}
\usepackage{marvosym}

\usepackage{multirow}

\usepackage[symbol]{footmisc}

\usepackage{pifont}
\newcommand{\xmark}{\ding{55}}%

\usepackage{amsmath}
\usepackage{amssymb}

\usepackage{caption}
\usepackage{subcaption}

\usepackage{acronym}
\acrodef{CMRI}{Cardiac Magnetic Resonance Imaging}
\acrodef{DDPM}{Denoising Diffusion Probabilistic Model}
\acrodef{DP}{Differential Privacy}
\acrodef{DL}{Deep Learning}
\acrodef{GAN}{Generative Adversarial Network}
\acrodef{LDM}{Latent Diffusion Model}
\acrodef{LV}{Left Ventricle} 
\acrodef{MRI}{Magnetic Resonance Imaging}
\acrodef{RV}{Right Ventricle} 
\acrodef{UKBB}{UK Biobank}
\acrodef{ACDC}[ACDC]{Automatic Cardiac Diagnosis Challenge}
\acrodef{AE}[AE]{AutoEncoder}
\acrodef{BMI}[BMI]{body mass index}
\acrodef{CMRI}[CMRI]{Cardiac Magnetic Resonance Imaging}
\acrodef{CFG}[CFG]{Classifier-Free Guidance}
\acrodef{CPA}[CPA]{Cardiac Phase Accuracy}
\acrodef{CSC}[CSC]{Cardiac Structure Correlation}
\acrodef{CT}[CT]{Computed Tomography}
\acrodef{DDPM}[DDPM]{Denoising Diffusion Probabilistic Model}
\acrodef{DL}[DL]{Deep Learning}
\acrodef{DP}[DP]{Differential Privacy}
\acrodef{DP-SGD}[DP-SGD]{Differentially Private Stochastic Gradient Descent}
\acrodef{ED}[ED]{end-diastolic}
\acrodef{ES}[ES]{end-systolic}
\acrodef{FID}[FID]{Fréchet Inception Distance}
\acrodef{FRD}[FRD]{Fréchet Radiomics Distance}
\acrodef{GAN}[GAN]{Generative Adversarial Network}
\acrodef{KL}[KL]{Kullback-Leibler}
\acrodef{LDM}[LDM]{Latent Diffusion Model}
\acrodef{LPIPS}[LPIPS]{Learned Perceptual Image Patch Similarity}
\acrodef{LV}{Left Ventricle} 
\acrodef{ML}[ML]{Machine Learning}
\acrodef{MLP}[MLP]{Multi-Layer Perceptron}
\acrodef{MRI}[MRI]{Magnetic Resonance Imaging}
\acrodef{MSE}[MSE]{Mean Squared Error}
\acrodef{MS-SSIM}[MS-SSIM]{Multi-Scale Structural Similarity Index}
\acrodef{PT}[PT]{pre-training}
\acrodef{ReLU}[ReLU]{Rectified Linear Unit}
\acrodef{RV}{Right Ventricle} 
\acrodef{SGD}[SGD]{Stochastic Gradient Descent}
\acrodef{SiLU}[SiLU]{Sigmoid Linear Unit}
\acrodef{SSIM}[SSIM]{Structural Similarity Index}
\acrodef{UKBB}{UK Biobank}
\acrodef{VAE}[VAE]{Variational AutoEncoder}
\acrodef{VQ-VAE}[VQ-VAE]{Vector Quantized-Variational AutoEncoder}
\acrodef{WRN}[WRN]{Wide Residual Network}
\begin{document}
\title{On Differentially Private 3D Medical Image Synthesis with Controllable Latent Diffusion Models}
\titlerunning{On Differentially Private 3D Medical Image Synthesis}
%\author{First Author\inst{1}\orcidID{0000-1111-2222-3333} \and
%Second Author\inst{2,3}\orcidID{1111-2222-3333-4444} \and
%Third Author\inst{3}\orcidID{2222--3333-4444-5555}}

\author{Deniz Daum \inst{1,2} %index{Daum Deniz}
\and Richard Osuala \inst{1,2,3} \orcidlink{0000-0003-1835-8564} %index{Osuala Richard}
\and Anneliese Riess \inst{1,2,4} \orcidlink{0009-0009-7087-7864} %{Riess Anneliese}
\and Georgios Kaissis \inst{1,2,5}\orcidlink{0000-0001-8382-8062}  %index{Kaissis, Georgios} 
\and Julia A. Schnabel \inst{1,2,6}\textsuperscript{*}\orcidlink{0000-0001-6107-3009} %index{Schnabel, Julia A.}
\and Maxime Di Folco \textsuperscript{(\Letter)}\inst{1,2}\textsuperscript{*} \orcidlink{0000-0001-6160-8050}} %index{Di Folco, Maxime}

\authorrunning{D. Daum et al.}

%\institute{Princeton University, Princeton NJ 08544, USA \and
%Springer Heidelberg, Tiergartenstr. 17, 69121 Heidelberg, Germany
%\email***@*****.**}

\institute{Technical University of Munich, Germany 
 \and Institute of Machine Learning in Biomedical Imaging, Helmholtz  Munich, Germany
 \and
Departament de Matemàtiques i Informàtica, Universitat de Barcelona, Spain \and
  Helmholtz AI, Helmholtz  Munich, Germany 
\and
Department of Computing, Imperial College London, UK \and
School of Biomedical Engineering \& Imaging Sciences, King's College London, UK\\
\email{maxime.difolco@helmholtz-munich.de}}

\maketitle              % typeset the header of the contribution
\begin{abstract}

Generally, the small size of public medical imaging datasets coupled with stringent privacy concerns, hampers the advancement of data-hungry deep learning models in medical imaging. This study addresses these challenges for 3D cardiac MRI images in the short-axis view. We propose Latent Diffusion Models that generate synthetic images conditioned on medical attributes, while ensuring patient privacy through differentially private model training. To our knowledge, this is the first work to apply and quantify differential privacy in 3D medical image generation. We pre-train our models on public data and finetune them with differential privacy on the UK Biobank dataset. Our experiments reveal that pre-training significantly improves model performance, achieving a Fréchet Inception Distance (FID) of 26.77 at $\epsilon=10$, compared to 92.52 for models without pre-training. Additionally, we explore the trade-off between privacy constraints and image quality, investigating how tighter privacy budgets affect output controllability and may lead to degraded performance. Our results demonstrate that proper consideration during training with differential privacy can substantially improve the quality of synthetic cardiac MRI images, but there are still notable challenges in achieving consistent medical realism.
%Our results demonstrate that high-quality synthetic cardiac MRI images can be generated while preserving privacy, offering a promising solution to augment limited datasets in medical imaging.
%We share our accessible codebase and model weights at https://github.com/anonymized_url}.
Code: \url{https://github.com/compai-lab/2024-miccai-dgm-daum}

%The abstract should briefly summarize the contents of the paper in

\end{abstract}

\input{sections/01_Introduction}

%\input{sections/02_Background}
\input{sections/03_Methods}
\input{sections/04_Experiments_Results}

\input{sections/05_Discussion_Conclusion}

\begin{credits}
\subsubsection{\ackname} This research has been conducted using the UK Biobank Resource under Application Number 87065. GK received support from the German Federal Ministry of Education and Research and the Bavarian State Ministry for Science and the Arts under the Munich Centre for Machine Learning (MCML), from the German Ministry of Education and Research and the the Medical Informatics Initiative as part of the PrivateAIM Project, from the Bavarian Collaborative Research Project PRIPREKI of the Free State of Bavaria Funding Programme "Artificial Intelligence -- Data Science", and from the German Academic Exchange Service (DAAD) under the Kondrad Zuse School of Excellence for Reliable AI (RelAI). RO acknowledges funding from the European Union’s Horizon research and innovation programmes under grant agreements No 952103 (EuCanImage), No 101057699 (RadioVal), and a research stay grant from the Helmholtz Information and Data Science Academy (HIDA).

\subsubsection{\discintname}
The authors have no competing interests to declare that are relevant to the content of this article. 

% \subsubsection{\discintname}
% It is now necessary to declare any competing interests or to specifically
% state that the authors have no competing interests. Please place the
% statement with a bold run-in heading in small font size beneath the
% (optional) acknowledgments\footnote{If EquinOCS, our proceedings submission
% system, is used, then the disclaimer can be provided directly in the system.},
% for example: The authors have no competing interests to declare that are
% relevant to the content of this article. Or: Author A has received research
% grants from Company W. Author B has received a speaker honorarium from
% Company X and owns stock in Company Y. Author C is a member of committee Z.
\end{credits}
\bibliographystyle{splncs04}
\bibliography{Paper-0019}

%\appendix
%\include{sections/Appendix}

\end{document}

%% file: sections/01_Introduction.tex
% !TeX root = ../main.tex

\section{Introduction}
\let\thefootnote\relax\footnotetext{*Equal contribution}
% Add UKBB
% MIMIC dataset
% 
The scarcity of public and large datasets in medical imaging poses a significant hurdle to training \ac{DL} models, as these extensive collections are crucial for developing accurate and robust models across various domains. One of the primary challenges is the privacy concerns associated with sharing medical data, leading to the availability of generally only small public datasets or larger ones with restricted access (such as \ac{UKBB} \cite{Bycroft2018TheData}), hampering the development and generalization of advanced medical imaging techniques \cite{Muller-Franzes2023ASynthesis}. As a solution, generating synthetic data from private datasets, with privacy guarantees, presents a promising approach. This strategy, especially when combined with traditional data augmentation techniques, can help overcome the limitations posed by the lack of large public datasets \cite{Pinaya2022BrainModels}. In recent years, various generative models have been employed to mitigate data scarcity in medical imaging. Among these, \acp{GAN} \cite{Goodfellow2014GenerativeNets} have gained popularity for generating synthetic data in different medical domains, such as cardiac \cite{Diller2020UtilityDisease} and brain \cite{Alrashedy2022BrainGAN:Models} \ac{MRI}. However, despite their wide adoption, \acp{GAN} face severe challenges including the failure to capture true diversity and unstable training dynamics, necessitating alternatives \cite{Muller-Franzes2023ASynthesis}. \acp{DDPM} have shown promise in learning the data distribution of medical images, showing success in areas like 3D brain MRI generation \cite{Dorjsembe2022Three-DimensionalModels}. Furthermore, \acp{LDM} have demonstrated their effectiveness in producing images that maintain consistent quality when conditioned on dense vectors \cite{Rombach2022High-ResolutionModels}, which can be leveraged to control the generation process for medical images using selected confounding variables \cite{Pinaya2022BrainModels}. While synthetic image generation using diffusion models has been explored for various applications (e.g., brain, lungs, knee, chest) and modalities (e.g., MRI, CT) \cite{Kazerouni2023DiffusionSurvey}, they remain underexplored in the cardiac domain \cite{Skorupko2024DebiasingModels}, where \acp{DDPM} have yet to be applied to 3D cardiac imaging. This gap is particularly problematic given the generally small size of public datasets for 3D cardiac imaging, which commonly range from tens to a few hundred patients \cite{Jafari2023AutomatedReview}.

Furthermore, although synthetic data may mitigate the limitations of small datasets in the medical domain and enhance privacy, only few studies have objectively addressed the privacy concerns associated with synthetic data. Research has explored privacy-preserving sampling methods post-model training \cite{Packhauser2023GenerationSystems} and local differential privacy, which introduces noise to images before model training \cite{Shibata2023LocalModels}. However, privacy-preserving sampling does not make the model inherently private, preventing its use for sharing the image distribution. Local differential privacy requires high noise to achieve reasonable privacy guarantees, often compromising the performance of the model. In contrast, differential privacy, which adds noise during the training process, can achieve higher accuracy under the same privacy guarantees \cite{Bebensee2019LocalTutorial}. Nevertheless, generative models inherently trained with differential privacy and their performance in medical applications remain underexplored. While Ghalebikesab et al. \cite{Ghalebikesabi2023DifferentiallyImages} experimented with medical image synthesis, their primary focus was on non-medical 2D CIFAR-10 images, and did not assess the impact on conditioning performance or privacy settings in a medical context. Specifically, the integration of \ac{DP} with diffusion models for 3D medical image generation and the quantitative assessment of the variety and quality of images produced with and without privacy mechanisms remain unexplored. Similarly, the impact of differentially private training on the conditioning mechanism has yet to be investigated.

In this study, we aim to address these gaps with four key contributions. (1) To the best of our knowledge, we introduce the first differentially private \ac{DDPM} for 3D medical image synthesis. (2) We apply \acp{LDM} to 3D cardiac MRI for the first time, analyzing image quality and diversity. (3) We establish clinically relevant metrics to evaluate the conditioning capabilities of cardiac MRI synthesis. (4) Our ablation study on privacy budgets shows that pre-training on a public dataset enables our \ac{DP} models to substantially improve the quality of the generated 3D cardiac MRI images.

%% file: sections/03_Methods.tex
% !TeX root = ../main.tex

\section{Methods}
\subsection{Latent Diffusion Model}

\acp{LDM} are a type of \acp{DDPM}, which are generative methods that reverse a noising process. \acp{DDPM} add noise to data over $T$ steps and learn to reverse this process, transforming noise back into the original data distribution \cite{Ho2020DenoisingModels}. \acp{LDM} enhance this process by using learned compression models to transform images into a latent representation with a \ac{VAE} \cite{Kingma2013Auto-EncodingBayes} or \ac{VQ-VAE} \cite{Oord2017NeuralLearning}, then training a diffusion model to denoise within the compressed data space. This method combines the generative power of \acp{DDPM} with the efficiency of working in a lower-dimensional space, resulting in faster training and sampling speeds while maintaining high image quality \cite{Rombach2022High-ResolutionModels}.

In this paper, we extend the standard 2D \ac{LDM} architecture proposed by \cite{Rombach2022High-ResolutionModels} to 3D by introducing 3D convolutions sized 1×3×3 as a replacement for the vanilla 2D convolutional layers and incorporating depth-wise self-attention layers into each transformer block at all resolution levels beyond the first, similar to \cite{Khader2023DenoisingGeneration}. We enhance depth handling further by introducing sinusoidal depth embeddings to the depth-wise self-attention layers and further conditioned the generation process on medical attributes through cross-attention mechanisms within the transformer layers of the denoising U-Net \cite{Ronneberger2015U-Net:Segmentation,Vaswani2017AttentionNeed}. For this purpose, we encode each scalar context attribute into an \(E\)-dimensional vector using a dedicated linear layer for each attribute which are then passed through a self-attention mechanism with which we model the interrelationships among the context attributes.

\subsection{Differential Privacy}

\ac{DP} ensures the privacy of sensitive data while enabling analysis by adding noise to the output of a mechanism. The \ac{DP} version of SGD, \ac{DP-SGD}, limits the influence of individual data points by applying gradient clipping and noise during training \cite{Abadi2016DeepPrivacy}. Formally, a mechanism \(M\) is \((\epsilon, \delta)\)-differentially private if for any datasets \(D, D'\) differing by one sample, and any subset of outcomes \(R\):
\[
P(M(D) \in R) \leq e^{\epsilon} \times P(M(D') \in R) + \delta 
\]
\cite{Abadi2016DeepPrivacy}.
Here, \(\epsilon\) indicates the privacy assurance level, and \(\delta\) the probability of privacy failure. Lower values for \(\epsilon\) and \(\delta\) imply higher privacy, which often comes at the cost of reduced utility of the mechanism.

Our \ac{LDM} architecture comprises an image-to-latent compression model and a latent space noise predicting network. Training both with \ac{DP} results in a composition, where the privacy guarantees accumulate. Hence, to achieve a certain privacy level \(\epsilon_{\text{target}}\), both the compressive model and the denoising network must be trained with stricter privacy guarantees. This would necessitate stronger privacy constraints for each model, potentially leading to a significant loss in model performance due to increased noise.

Conversely, if the compressive model uses only public data, we avoid this complication. The privacy level of the denoising network alone then dictates the privacy level for the entire mechanism. Therefore, if the compression model can perform well without using the private dataset, it is advisable to avoid \ac{DP} training for the compressive model, allowing higher privacy levels to be maintained. Hence, we chose to train only the denoising network with \ac{DP} and the compressive model on public data, leveraging Opacus \cite{Yousefpour2021Opacus:PyTorch} for \ac{DP-SGD} implementation.

\subsection{Data and Pre-processing}

We processed the short-axis \ac{CMRI} images from the \ac{UKBB} (over 52,000 patient exams) at \ac{ES} and \ac{ED} phases. The dataset was randomly split per patient into training ($N=42,192$), validation ($N=5,274$), and testing ($N=5,274$) sets with an 80-10-10 ratio. Pre-processing included resampling (1.8269 $\times$ 1.8269 $\times$ 10 mm), normalization of pixel values (0 to 1), rotation and centering based on segmentations calculated using the \textit{ukbb\_cardiac} package \cite{Bai2018AutomatedNetworks}, center cropping ($96 \times 96$) around the cardiac region, and padding to 13 slices with mean padding. We pre-trained \ac{DP} models with the public \ac{ACDC} dataset (150 patients, 4D cine MRI), applying similar pre-processing steps. We applied data augmentation included scaling, rotating, and shifting during training. Pre-training was done in two steps. First we pre-trained the \ac{LDM} on each frame of the 3D images of the cine \ac{MRI} sequence of the \ac{ACDC} dataset without any context attributes. In another step, we then trained only on images in \ac{ES} and \ac{ED} phases for which we also used context attributes.

A list of the selected attributes for conditioning the generation process can be found in the supplementary material. As the attribute \textit{Sex} is not provided in the \ac{ACDC} dataset, we assign its value randomly during pre-training.

\subsection{Evaluation Metrics}
We assessed the realism of the generated images using the \ac{FID} \cite{Heusel2017GANsEquilibrium} and its diversity using the \ac{MS-SSIM} \cite{Wang2003MultiscaleAssessment}. Both metrics are calculated using only the central seven slices to avoid the influence of padded dimensions.

Inspired by the conditioning analysis of \cite{Pinaya2022BrainModels} we propose cardiac imaging specific metrics for conditional synthetic data evaluation. We measure the accuracy of generated images being classified into their intended phase (end-diastolic or end-systolic). For that purpose, we trained a model on the \ac{UKBB} dataset to classify these phases, achieving an accuracy of 99.9\%. The percentage of correctly classified generated images is referred to as \ac{CPA}.

Additionally, we evaluate how key cardiac attributes used for conditioning, such as left and right ventricular volumes and myocardial wall thickness, correlate with measurements in generated images. These values are computed using the \textit{ukbb\_cardiac} segmentation network trained on the \ac{UKBB} dataset, which achieves a Dice score of 89.18\%. The correlation coefficients between these values and their designated context labels are termed \ac{CSC} metrics, specifically LV-\ac{CSC}, RV-\ac{CSC}, and MYO-\ac{CSC}.

%% file: sections/04_Experiments_Results.tex
% !TeX root = ../main.tex

\section{Experiments \& Results}
\subsection{Implementation Details}
We used the same neural backbone for all experiments, differing only in datasets, input sizes, and training regimes (Adam optimizer or \ac{DP-SGD}). The noise-predicting U-Net comprised 3 resolution levels (channels 128, 256, 384), each with 2 residual blocks, and transformers in the lower levels with 8 attention heads (dimension 16). Context attributes were embedded to \(E=16\). A fixed linear variance schedule was used with \(\beta_{\min} = 1.0 \times 10^{-4}\), \(\beta_{\max} = 7.0 \times 10^{-2}\), and 150 denoising steps. The learning rate started at \(2.0 \times 10^{-5}\) and was reduced by 10\% every 25 epochs. 

As image-to-latent compression model, we conducted our experiments using both a $\beta$-\ac{VAE} with \(\beta=0.1\) and a \ac{VQ-VAE} with a codebook size of 4000 and a dimension of 4. Each model exhibited two resolution levels with channels of 64 and 128 and compressed the 3D images from a size of \(13 \times 96 \times 96\) to \(13 \times 24 \times 24\). For the base models without privacy considerations the compressive stage was trained on the \ac{UKBB} and for \ac{DP} models on the public \ac{ACDC} dataset.

Base \acp{LDM} were trained on the \ac{UKBB} dataset for 150 epochs with a batch size of 64. For \ac{DP} \acp{LDM}, we tested a naive approach by training the denoising process from scratch with \ac{DP-SGD} for 150 epochs with a batch size of 256. The focus of this study, however, is on the pre-trained \ac{DP} \acp{LDM}. These models were initially pre-trained on the \ac{ACDC} dataset for 150 epochs on all frames of the cine \ac{MRI} sequence without context attributes, followed by 75 epochs on images in the \ac{ES} and \ac{ED} phases with context attributes, both with a batch size of 64. The pre-trained models were then finetuned on the \ac{UKBB} dataset for 75 epochs with \ac{DP-SGD} at privacy levels \(\epsilon \in [0.1, 1, 10]\), using a batch size of 256.

Training the base \acp{LDM} and pre-training was performed using two A100-80GB GPUs, taking about one day to complete. The naive \ac{DP} \acp{LDM} training and the \ac{DP} finetuning were conducted using eight Tesla V100-32GB GPUs, with the former finishing within $\approx$20 hours and the latter within $\approx$12 hours.

To improve pre-trained \ac{DP} model performance, we used \ac{CFG} \cite{Ho2022Classifier-FreeGuidance} as a post-training technique, applying guidance parameters \(G \in [1, 3, 7]\) for $\epsilon=10$ models to analyze trade-offs between generation quality and conditioning performance.

\subsection{Results}

% MS-SSIM delta??

\subsubsection{Generation results}

\begin{table}[t]
\centering
\caption{Performance comparison of $\beta$-VAE and VQ-VAE models with different differential privacy settings. A MS-SSIM of 0.196 is the baseline for real images.}
\label{table:ddpm_results}
\begin{tabular}{m{0.12\linewidth}m{0.06\linewidth}m{0.06\linewidth}m{0.06\linewidth}m{0.01\linewidth}m{0.08\linewidth}m{0.12\textwidth}m{0.1\linewidth}m{0.1\linewidth}m{0.1\linewidth}>{\centering\arraybackslash}m{0.1\linewidth}}
\toprule
\multirow{2}{*}{\textbf{Model}} & \multirow{2}{*}{\textbf{DP}} &  \multirow{2}{*}{\textbf{PT}} & \multirow{2}{*}{\textcolor{white}{as}\boldmath{$\epsilon$}} & & \multirow{2}{*}{\textbf{FID} $\downarrow$} &  \textbf{MS-} & \multicolumn{3}{c}{\textbf{CSC} $\uparrow$} &  \multirow{2}{*}{\textbf{CPA $\uparrow$}}  \\
\cmidrule(lr){8-10}
& & & & & & \textbf{SSIM} & \textbf{LV} & \textbf{RV} & \textbf{MYO} & \\
\midrule
\multirow{5}{*}{{$\beta$-VAE}} & \multicolumn{1}{|c}{\xmark}  & \multicolumn{1}{c}{\xmark} &  \multicolumn{1}{c|}{} & & \textbf{15.36} & 0.223 & \textbf{0.95} & \textbf{0.95} & \textbf{0.94} & \textbf{1.00} \\ %2024_05_12_15_10_11_814081
%\cmidrule(lr){2-10}

 & \multicolumn{1}{|c}{\checkmark} & \multicolumn{1}{c}{\xmark} & \multicolumn{1}{c|}{10} & & 92.52 & 0.153 & 0.05 & 0.05 & 0.31 & 0.50 \\ %2024_05_27_11_41_02_304953 
 %\cmidrule(lr){2-4}
 & \multicolumn{1}{|c}{\checkmark} & \multicolumn{1}{c}{\checkmark} & \multicolumn{1}{c|}{10} & & 26.77 & 0.196 & 0.21 & 0.19 & 0.02 & 0.53 \\ %2024_05_25_08_25_17_213700
 & \multicolumn{1}{|c}{\checkmark} & \multicolumn{1}{c}{\checkmark} & \multicolumn{1}{c|}{1}& & 29.68 & 0.183 & 0.14 & 0.13 & 0.03 & 0.53 \\ %2024_05_28_10_47_40_224423
 & \multicolumn{1}{|c}{\checkmark} & \multicolumn{1}{c}{\checkmark} & \multicolumn{1}{c|}{0.1}& & 42.67 & 0.143 & 0.17 & 0.13 & 0.03 & 0.52 \\ %2024_05_29_08_53_49_443683
\midrule
\multirow{5}{*}{{VQ-VAE}} & \multicolumn{1}{|c}{\xmark} & \multicolumn{1}{c}{\xmark} & \multicolumn{1}{c|}{} & & 16.27 & 0.233 & \textbf{0.95} & \textbf{0.95} & 0.93 & \textbf{1.00} \\ %2024_05_12_15_15_30_247298
 & \multicolumn{1}{|c}{\checkmark} & \multicolumn{1}{c}{\xmark} & \multicolumn{1}{c|}{10}& & 75.18 & 0.193 & 0.05 & 0.01 & 0.26 & 0.52 \\ %2024_06_07_22_09_20_299970
 & \multicolumn{1}{|c}{\checkmark} & \multicolumn{1}{c}{\checkmark} & \multicolumn{1}{c|}{10}& & 31.10 & 0.228 & 0.15 & 0.14 & 0.05 & 0.52 \\ %2024_05_30_13_01_07_180596
 & \multicolumn{1}{|c}{\checkmark} & \multicolumn{1}{c}{\checkmark} & \multicolumn{1}{c|}{1} && 32.96 & 0.204 & 0.11 & 0.11 & 0.01 & 0.52 \\ %2024_06_01_00_13_47_656641
 & \multicolumn{1}{|c}{\checkmark} & \multicolumn{1}{c}{\checkmark} & \multicolumn{1}{c|}{0.1} && 39.25 & 0.163 & 0.11 & 0.09 & 0.02 & 0.51 \\  %2024_06_01_10_51_55_894129
\bottomrule
\end{tabular}
\end{table}

Base models generated convincing \ac{CMRI} images as illustrated in \autoref{fig:GeneratedImages} for $\beta$-VAE. The $\beta$-VAE achieved superior results with a lower \ac{FID} of 15.36 compared to \ac{VQ-VAE}'s 16.27 and higher diversity (\ac{MS-SSIM} of 0.223 vs. 0.233), while real images had a baseline diversity of 0.196.  Similarly, conditioning performance yielded high correlation scores as shown in \autoref{table:ddpm_results}. \ac{DP} models without \ac{PT} failed in generating high-quality images (\ac{FID} of 92.52 and 75.18), where \ac{VQ-VAE} produced slightly more recognizable cardiac structures compared to the nearly unrecognizable structures in $\beta$-VAE. Pre-training substantially improved results, achieving \ac{FID} of 26.77 and 31.10 at $\epsilon=10$. The $\beta$-VAE model achieved a \ac{MS-SSIM} of 0.196. Conditioning performance (LV-/RV-\ac{CSC}, MYO-\ac{CSC}, \ac{CPA}) showed low correlation scores, indicating minimal learned concepts.

\subsubsection{Privacy trade-off}

\begin{figure}[t]
    \centering
    \includegraphics[width=\linewidth]{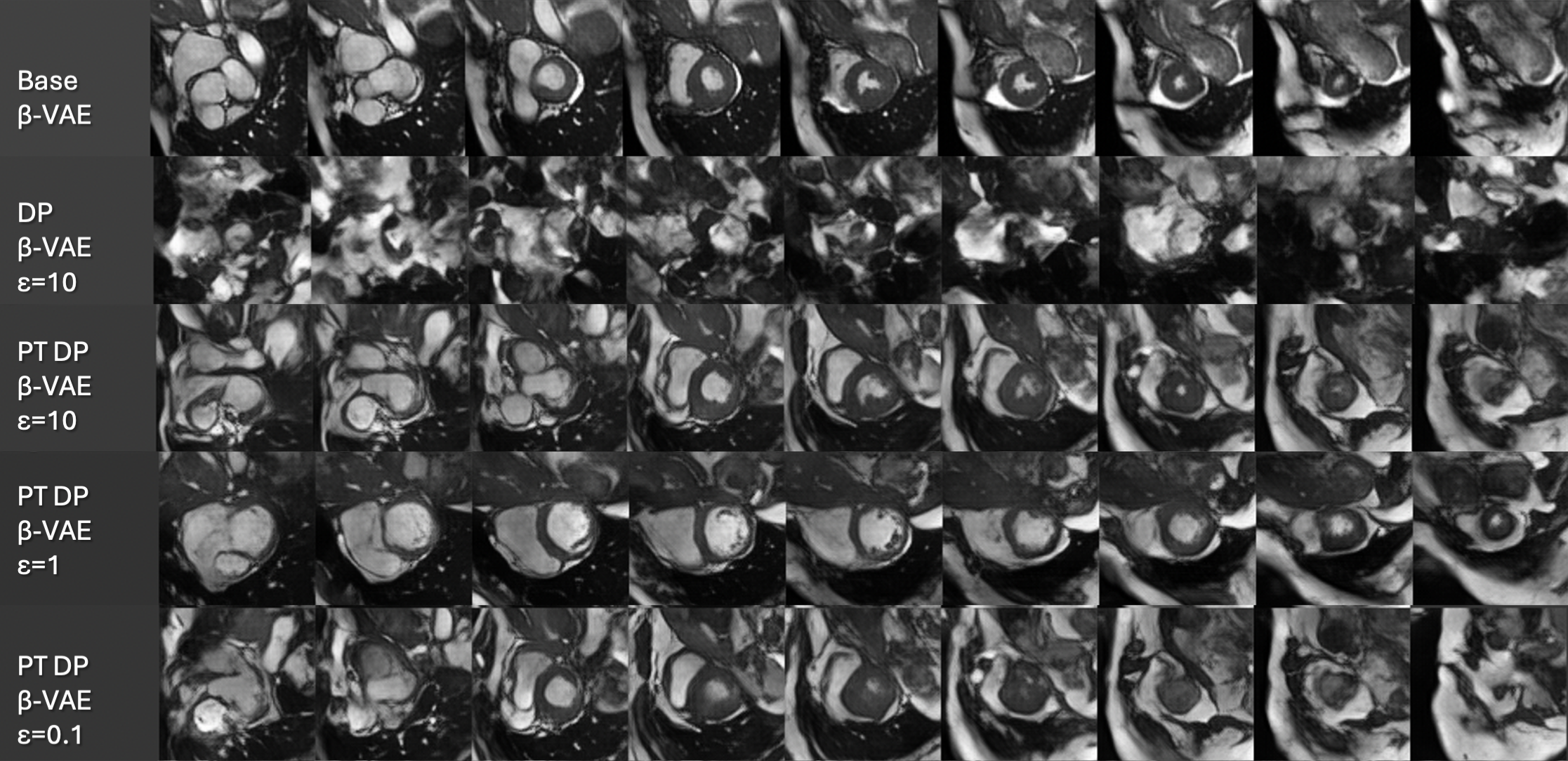}
    \caption{Synthetic 3D CMRI images at ES phase from $\beta$-VAE models with and without DP, PT and various privacy levels. From left to right: slices from the basal level to the apex.}
    \label{fig:GeneratedImages}
\end{figure}

\begin{figure}
    \centering
    \includegraphics[width=1\linewidth]{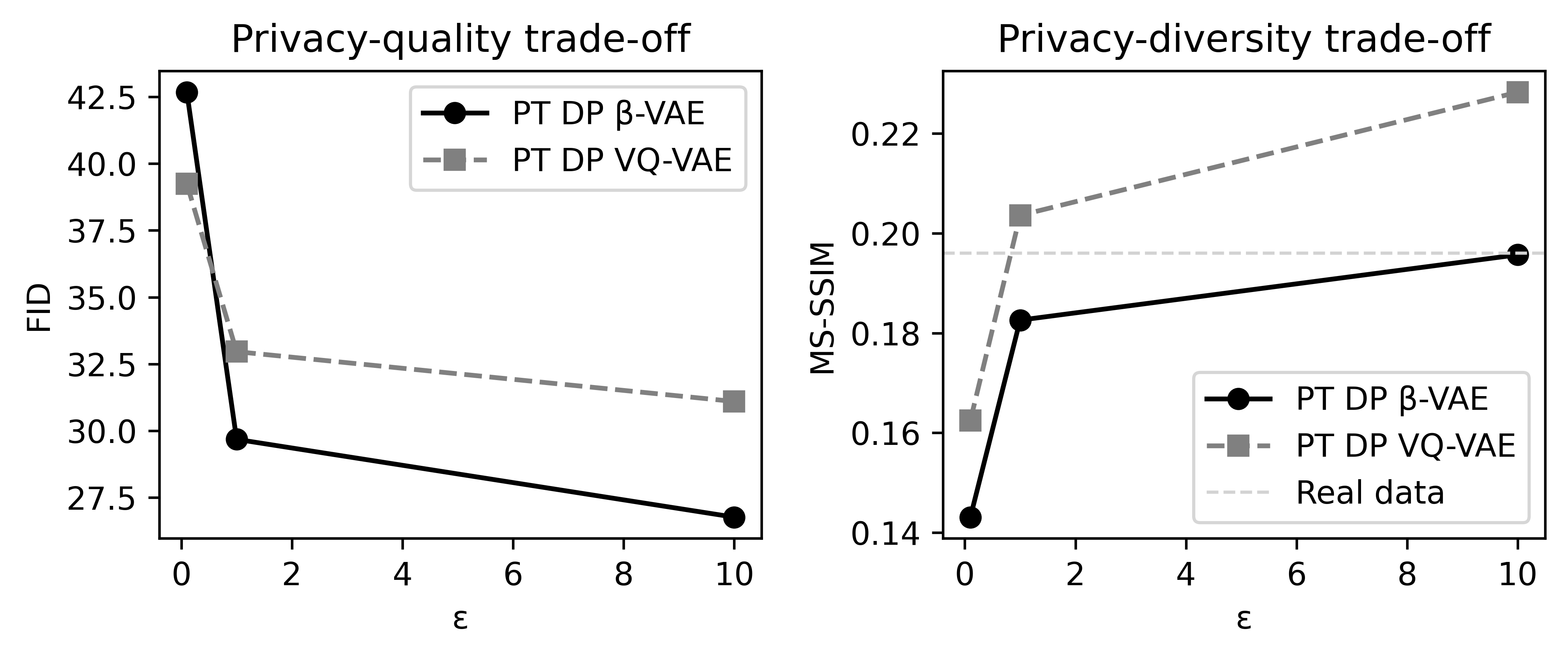}
    \caption{Trade-off between generation quality and various privacy levels.}
    \label{fig:privacy_tradeoff}
\end{figure}

Changing the privacy level showed a clear trade-off between privacy and image quality. For the $\beta$-VAE model, reducing the privacy budget from \(\epsilon=10\) to \(\epsilon=0.1\) significantly degraded image quality (\ac{FID} from 26.77 to 42.67) and conditioning scores. The \ac{VQ-VAE} model exhibited a similar decline (\ac{FID} from 31.10 to 39.25), with \ac{MS-SSIM} scores also dropping below the benchmark of 0.196, indicating less coherence and more out-of-distribution shapes at higher privacy levels. This trade-off is depicted in \autoref{fig:privacy_tradeoff}, illustrating how quality and coherence diminish as privacy increases. A qualitative assessment of $\beta$-VAE images in \autoref{fig:GeneratedImages} supported these observations. Images generated at \(\epsilon=10\) displayed better-defined cardiac structures compared to blurred and inconsistent images at \(\epsilon=0.1\), corroborating the lower \ac{MS-SSIM} and higher \ac{FID} values. The same illustration for VQ-VAE is shown in the supplementary materials.

\subsubsection{Conditioning trade-off}

\begin{figure}[!htb]
    \centering
    \begin{minipage}{0.56\linewidth}
        \centering
        \begin{tabular}{llccccccc}
        \toprule
         \multirow{2}{*}{\textbf{G}} & \textcolor{white}{a}&\multirow{2}{*}{\textbf{FID} $\downarrow$} & \textbf{MS} & \multicolumn{3}{c}{\textbf{CSC} $\uparrow$} & \multirow{2}{*}{\textbf{CPA} $\uparrow$} \\
        \cmidrule(lr){5-7}
         & & & \textbf{-SSIM} & \textbf{LV} & \textbf{RV} & \textbf{MYO} & \\
        \midrule
         \multicolumn{1}{l|}{\(0\)} & & \textbf{26.77} & 0.20 & 0.21 & 0.19 & 0.02 & 0.53 \\ %2024_05_25_08_25_17_213700
        \multicolumn{1}{l|}{\(1\)} & & 26.82 & 0.18 & 0.34 & 0.31 & \textbf{0.03} & 0.54 \\ %2024_05_27_11_43_15_767801
        \multicolumn{1}{l|}{\(3\)} & & 29.32 & 0.16 & 0.47 & 0.45 & 0.00 & 0.57 \\ %2024_05_27_11_46_11_162399
        \multicolumn{1}{l|}{\(7\)} & & 39.38 & 0.14 & \textbf{0.53} & \textbf{0.51} & 0.02 & \textbf{0.57} \\ %2024_05_26_10_05_35_302519
        \bottomrule
        \end{tabular}
        \subcaption{CFG performance comparison of different guidance values.}
        \label{table:cfg_results}
    \end{minipage}%
    \hfill
    \begin{minipage}{0.4\linewidth}
        \centering
        \includegraphics[width=\linewidth]{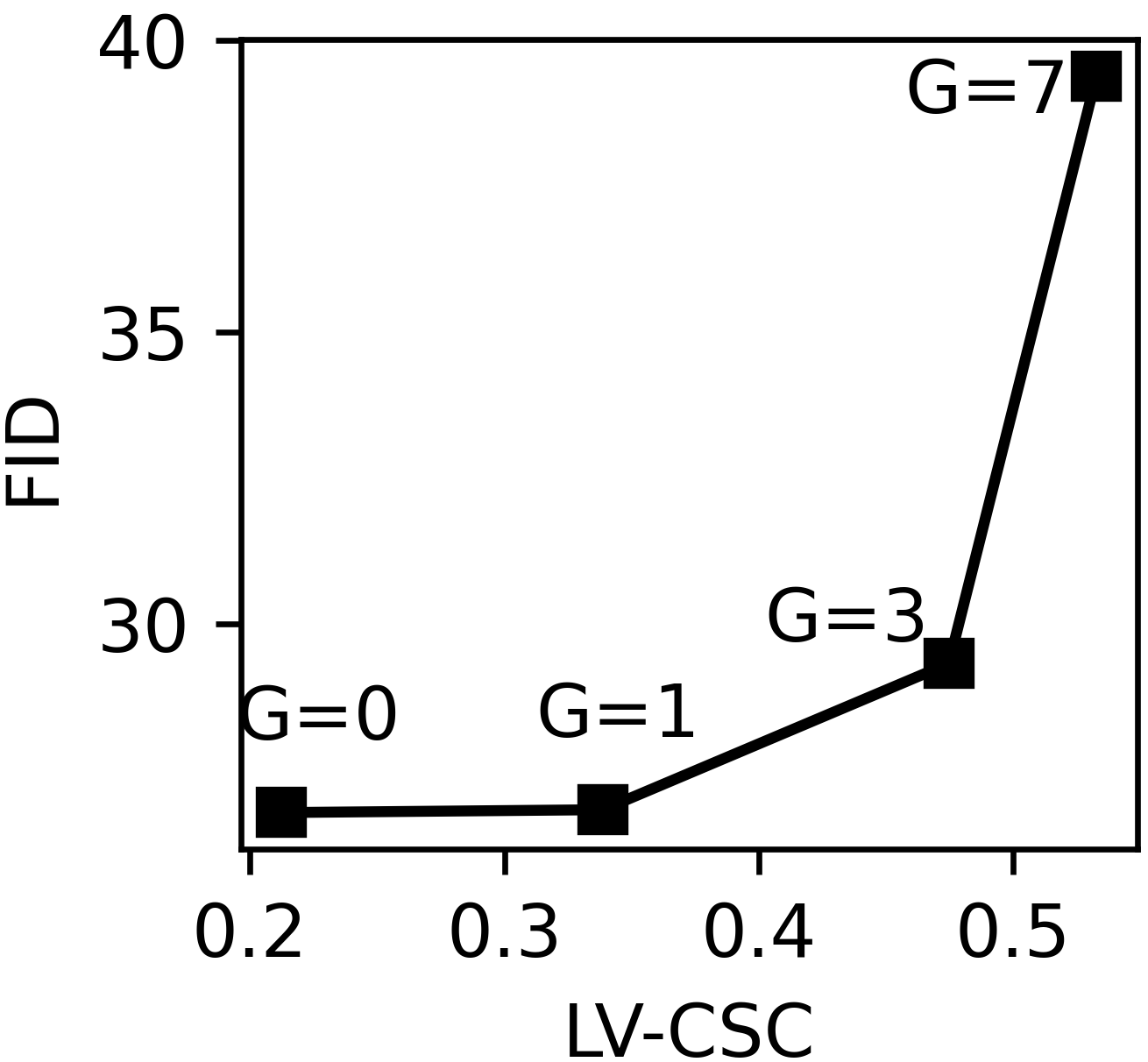}
        \subcaption{CFG trade-off between fidelity and conditioning performance.}
        \label{fig:cfg_tradeoff}
    \end{minipage}
    \caption{CFG performance comparison and trade-off for PT DP $\beta$-VAE at $\epsilon=10$.}
\end{figure}

Using \ac{CFG}, we investigated the trade-off between image quality and conditioning quality as shown in \autoref{table:cfg_results} for a privacy regime $\epsilon = 10$. Low to moderate guidance values ($G\in[1, 3]$) offered a sensible trade-off, maintaining low \ac{FID} while improving \ac{CSC}. Extreme guidance ($G=7$) improved conditioning but degraded image quality considerably. This trade-off is illustrated in \autoref{fig:cfg_tradeoff}, indicating that low to moderate guidance results in only minor quality degradation, which increases with higher \(G\). MYO-CSC and \ac{CPA} remained minimal across all \ac{DP} models.

% \begin{figure}
%   \begin{minipage}[b]{.45\linewidth}
%     \centering
%     \includegraphics[width=0.7\linewidth]{example-image}
%     \captionof{figure}{Figure caption}% \caption{Figure caption}
%   \end{minipage}\hfill
%   \begin{minipage}[b]{.45\linewidth}
%     \centering
%     \begin{tabular}{ *{5}{c} }
%       A & b & C & d & e \\
%       \hline
%       1 & 2 & 3 & 4 & 5
%     \end{tabular}
%     \captionof{table}{Table caption}
%   \end{minipage}
% \end{figure}

% \begin{figure}
% \begin{floatrow}

% \ffigbox{%
%   \\includegraphics[width=0.5\linewidth]{img/cfg_tradeoff.png}
% }
% \capbtabbox{%
%   \begin{tabular}{lccccccc}
% \toprule
% \textbf{Model} & \textbf{FID} $\downarrow$ & \textbf{MS-SSIM} & \multicolumn{3}{c}{\textbf{CSC} $\uparrow$} & \textbf{CPA} $\uparrow$ \\
% \cmidrule(lr){4-6}
% & & & \textbf{LV} & \textbf{RV} & \textbf{MYO} & \\
% \midrule
% \(G=0\) & \textbf{26.77} & 0.20 & 0.21 & 0.19 & 0.02 & 0.53 \\ %2024_05_25_08_25_17_213700
% \(G=1\) & 26.82 & 0.18 & 0.34 & 0.31 & \textbf{0.03} & 0.54 \\ %2024_05_27_11_43_15_767801
% \(G=3\) & 29.32 & 0.16 & 0.47 & 0.45 & 0.00 & 0.57 \\ %2024_05_27_11_46_11_162399
% \(G=7\) & 39.38 & 0.14 & \textbf{0.53} & \textbf{0.51} & 0.02 & \textbf{0.57} \\ %2024_05_26_10_05_35_302519
% \bottomrule
% \end{tabular}
% }

% \end{floatrow}
% \end{figure}

% subfigure

%% file: sections/05_Discussion_Conclusion.tex
% !TeX root = ../main.tex

\section{Discussion \& Conclusion}

% I have reorganised a bit the discussion

In this paper, we evaluated the impact of training with differential privacy for the synthesis of controllable 3D medical image synthesis. We showed, for 3D cardiac MRI, that without pre-training, \ac{DP} models produced unusable outputs while pre-training on a public dataset substantially improved \ac{DP} model performance, reducing \ac{FID} and enhancing key metrics. However, maintaining consistent medical realism is still difficult. Privacy trade-offs show that model performance degraded with stricter privacy budgets. Higher \(\epsilon\) values yield clearer images, while lower \(\epsilon\) values result in inconsistencies. The trade-off is not linear, with considerable quality deterioration below \(\epsilon=1\). A reasonable \(\epsilon\) range was found to be 1 and 10.

Additionally, we further explored \ac{CFG} demonstrating improvements in conditioning performance. Although \ac{DP} models showed notably lower conditioning performance compared to non-private models, they still managed to learn some concepts, such as the ventricular volumes. Using \ac{CFG} during inference allowed trading generation quality for conditioning accuracy. However, certain conditioning tasks (cardiac phase and myocardial wall thickness) remained challenging, likely due to differences in label distributions between pre-training and private datasets. Visual comparisons revealed that, despite lower fidelity (less texture variation), conditioning can still be achieved at higher guidance values. This is counterintuitive, as one would typically expect lower image fidelity to negatively impact conditioning performance, making the images more challenging to evaluate through the segmentation network, warranting further research.  

For the compressive stage, $\beta$-\ac{VAE} consistently outperformed \ac{VQ-VAE} in reconstruction performance. Experiments with \ac{DP} in the compressive stage revealed that \ac{VQ-VAE} had notable issues, resulting in blurred reconstructions, while $\beta$-\ac{VAE} maintained better image quality. However, training on public data yielded the best privacy-preserving performance, even with a small dataset.